%% file: cherenkov.tex
\documentclass[12pt]{iopart}

\usepackage[dvips]{graphicx}
\usepackage{fancyhdr}
\begin{document}


\title{Visualisation of \v{C}erenkov Radiation and the Fields of a Moving Charge}

\author{Robert N C Pfeifer$^1$ and Timo A Nieminen}

\address{School of Physical Sciences, The University of Queensland, Brisbane, QLD 4072, Australia}
\ead{$^1$pfeifer@physics.uq.edu.au}

\begin{abstract}
For some physics students, the concept of a particle travelling faster than the speed of light holds endless fascination, and \v{C}erenkov radiation is a visible consequence of a charged particle travelling through a medium at locally superluminal velocities. The Heaviside--Feynman equations for calculating the magnetic and electric fields of a moving charge have been known for many decades, but it is only recently that the computing power to plot the fields of such a particle has become readily available for student use. This article investigates and illustrates the calculation of Maxwell's $\bi{D}$ field in homogeneous isotropic media for arbitrary, including superluminal, constant velocity, and uses the results as a basis for discussing energy transfer in the electromagnetic field.
\end{abstract}
\pacs{41.60.-m, 41.60.Bq}
\medskip\medskip
\medskip\medskip
\noindent Published in \emph{European Journal of Physics} vol. 27 (2006) pp. 521-529
\maketitle

\section{Introduction}
For any medium, the speed of light is also the speed of propagation of electromagnetic waves within that medium --- after all, light is simply an electromagnetic wave. But what is the relationship between an electromagnetic wave and the electromagnetic field of a particle?


In answer to this, we consider the field of a stationary charge in free space.
The field of the charge $q$ extends to infinity and at distance $r$ is simply given by
\begin{equation}
\bi{E}=\frac{1}{4\pi\varepsilon_0}\frac{q}{r^2}\,\hat\bi{r}.
\end{equation}
If, however, we consider the field to be made up of an imaginary flux originating at the particle itself and travelling radially outwards with velocity $c$, we see that the flux present at distance $r$ at time $t$ will have originated at the particle at a time
\begin{equation}
t^\prime=t-\frac{r}{c}.
\end{equation}
When the particle is stationary, it is unnecessary to make this distinction as the point from which the field was emitted is the same as the point currently occupied by the particle. When the particle is in motion, however, it becomes necessary to identify both when and where the particle may have been located so as to give rise to the field at a particular point.

Let the particle follow an arbitrary path $\bi{x}(t)$. We may write the four-dimensional co--ordinates of any point on this path in the form $(\bi{x}(t),t)$. We now wish to determine the field at point $(\bi{x^\prime},t^\prime)$, which may or may not lie upon the path. In order for the field to have propagated here at speed $c$, it must have originated at a co--ordinate $(\bi{x},t)$ such that
\begin{equation} \label{eq:retposgeneral}
\left| \bi{x^\prime}-\bi{x} \right| = c(t^\prime-t).
\end{equation}
A value of $(\bi{x}(t),t)$ satisfying \eref{eq:retposgeneral} is referred to as the \emph{retarded position} of the particle generating the field at $(\bi{x^\prime},t^\prime)$. For certain trajectories, such as superluminal motion, there may be multiple retarded positions contributing to the field at a given location. These are combined using the principle of linear superposition.

So how do we determine the fields arising from the particle at the retarded position? The particle is in motion, and may even be accelerating. Obviously the Coulomb field is not appropriate; the moving particle constitutes a current as well as an electric charge. Writing $\bi{R}=\bi{x^\prime}-\bi{x}$ and $R=\left|\bi{R}\right|$, the equation for the relevant magnetic field
\begin{equation}
\eqalign{
\bi{B}=\frac{\mu_0 q}{4\pi}
\left[ \left( \frac{\dot \bi{x}\times \hat\bi{R}}{\kappa^2 R^2}\right)_{\mathrm{ret}}
      +\frac{1}{c(R)_{\mathrm{ret}}} \frac{\partial}{\partial t}\left( \frac{\dot \bi{x}\times \hat\bi{R}}{\kappa} \right)_{\mathrm{ret}} \right] \\
\kappa=1-\frac{\dot \bi{x}\cdot \hat\bi{R}}{c}
}
\end{equation}
is attributed to Oliver Heaviside \cite[p. 436]{heaviside}, and the explicit equation for the electric field
\begin{equation} \label{eq:feynman}
\bi{E}=\frac{q}{4\pi\varepsilon_0}
\left[ \left( \frac{\hat \bi{R}}{R^2} \right)_{\mathrm{ret}}
      +\frac{(R)_{\mathrm{ret}}}{c} \frac{\partial}{\partial t} \left(\frac{\hat \bi{R}}{R^2} \right)_{\mathrm{ret}}
      +\frac{\partial^2}{c^2\partial t^2}\left(\hat \bi{R}\right)_{\mathrm{ret}} \right] ,
\end{equation}
although also originally developed by Heaviside \cite[p. 437]{heaviside}, is usually attributed to Richard Feynman \cite[p. II-21-1]{feynman}. They are related by
\begin{equation}
\bi{B}=\hat \bi{R} \times \bi{E}.
\end{equation}
The corresponding scalar and vector potentials are known as the Li\'enard--Wiechert potentials
\begin{eqnarray}
\phi(\bi{x^\prime},t)=\left[ \frac{1}{4\pi\varepsilon}
                             \frac{q}{\bi{R}\cdot(1-\dot \bi{x}/c)} \right]_{\mathrm{ret}} \\
\bi{A}(\bi{x^\prime},t)=\left[ \frac{\mu}{4\pi}
                        \frac{q\,\dot \bi{x}/c}{\bi{R}\cdot(1-\dot \bi{x}/c)} \right]_{\mathrm{ret}}
\end{eqnarray}
Derivations of the above expressions may be found in advanced electromagnetics textbooks such as Jackson \cite{jackson}.

Now let us extend our considerations to media other than vacuum. In material media, the values of $\varepsilon$, $\mu$ and refractive index $n$ will generally differ from those in vacuum, and the medium may also be anisotropic. Furthermore, when a charge is in relativistic motion the medium will appear to contract along the direction of travel, with consequences for $\varepsilon$ and $\mu$. Thus even a medium which is isotropic at rest will appear anisotropic to a moving charge.

Our calculations shall be performed in the rest frame of an isotropic medium. However, by working with Maxwell's $\bi{D}$ and $\bi{H}$ fields we take an approach more readily extensible to the cases of moving and anisotropic media.

Because
\begin{equation}
\bi{D}=\varepsilon\bi{E}
\end{equation}
and
\begin{equation}
\bi{B}=\mu\bi{H},
\end{equation}
$\varepsilon$ and $\mu$ are eliminated from our equations and only the value of the refractive index affects our results. This alters the speed of propagation of field modulations within the medium,




\begin{equation}
c^\prime=\frac{c}{n},
\end{equation}
and thus affects the solutions for the retarded positions. The equations for the $\bi{D}$ and $\bi{H}$ fields then become
\begin{eqnarray}
\bi{D}=\frac{q}{4\pi}   \label{dfield}
\left[ \left( \frac{\hat \bi{R}}{R^2} \right)_{\mathrm{ret}}
      +\frac{(R)_{\mathrm{ret}}}{c} \frac{\partial}{\partial t} \left(\frac{\hat \bi{R}}{R^2} \right)_{\mathrm{ret}}
      +\frac{\partial^2}{c^2\partial t^2}\left(\hat \bi{R}\right)_{\mathrm{ret}} \right]
\\
\eqalign{
\bi{H}=&\frac{q}{4\pi}
\left[ \left( \frac{\dot \bi{x}\times \hat\bi{R}}{\kappa^2 R^2}\right)_{\mathrm{ret}}
      +\frac{1}{c(R)_{\mathrm{ret}}} \frac{\partial}{\partial t}\left( \frac{\dot \bi{x}\times \hat\bi{R}}{\kappa} \right)_{\mathrm{ret}} \right] \\
&\kappa=1-\frac{\dot \bi{x}\cdot \hat\bi{R}}{c^\prime}.
}
\end{eqnarray}
Note that the $c$ associated with each $\frac{\partial}{\partial t}$ is unchanged as it arises not from the field propagation time, but as a constant relating our units of measurement in temporal and spatial dimensions. This is unaffected by a change of 
material medium.

\section{Calculation of Retarded co--ordinates}
Consider a particle travelling in a straight line parallel to the $y$--axis at constant velocity $\bi{v}$. We choose our co--ordinate system such that the direction of travel is along the $x$--axis and wish to calculate the $\bi{D}$ field magnitudes in the $xy$ plane. If the particle is at $y$ co--ordinate $y_0$ at time $t_0$ then its equation of motion is given by
\begin{equation} \label{motion1}
y=y_0+v(t-t_0).
\end{equation}

Suppose we wish to calculate the $\bi{D}$ field at spatial co--ordinate $(x^\prime,y^\prime,0)$ at time $t^\prime$. We denote particle co--ordinates with unprimed characters and field co--ordinates with primed characters. As can be seen from \fref{singlemediumret}, the following relationships hold:
\begin{figure}
\includegraphics{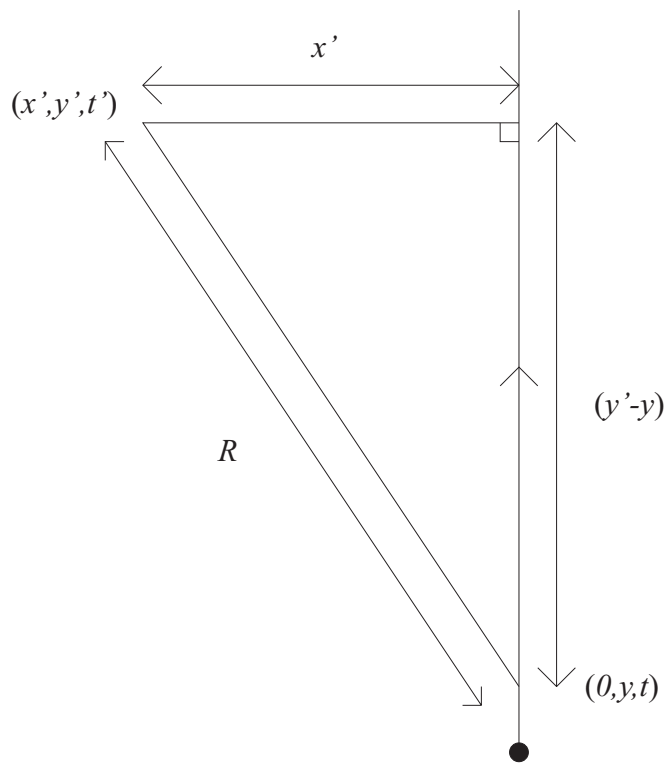}
\caption{Particle travelling through a homogeneous medium\label{singlemediumret}}
\end{figure}
\begin{eqnarray}
(t^\prime-t)=\frac{R}{c^\prime} = \frac{R\,n}{c} \label{r1}\\
R^2=x^{\prime\,2}+(y^\prime-y)^2 \label{r2}
\end{eqnarray}
Combining \eref{motion1}, \eref{r1} and \eref{r2}, and choosing $c=1$ and $t_0=0$ for clarity, we obtain a quadratic in $y$:
\begin{equation} \label{reteqn}
\eqalign{
y^2&(n^2v^2-1) + y(2y_0+2t^\prime v-2y^\prime n^2v^2) \\
 &+ (x^{\prime\,2}n^2v^2 + y^{\prime\,2}n^2v^2 -
 t^{\prime\,2}v^2 - y_0^2 - 2t^\prime y_0 v) = 0.
}
\end{equation}
In accordance with convention we discard solutions corresponding to advance potentials $(t>t^\prime)$, using \eref{motion1} to identify the time $t$ corresponding to a given source solution $y$. For subluminal particles
in a single medium, one of the solutions will be advanced and the other retarded.
For superluminal particles, solutions will either be both advanced, both retarded, or both imaginary, indicating that the field has not yet reached this region.

\begin{figure}
\includegraphics[width=0.9\columnwidth]{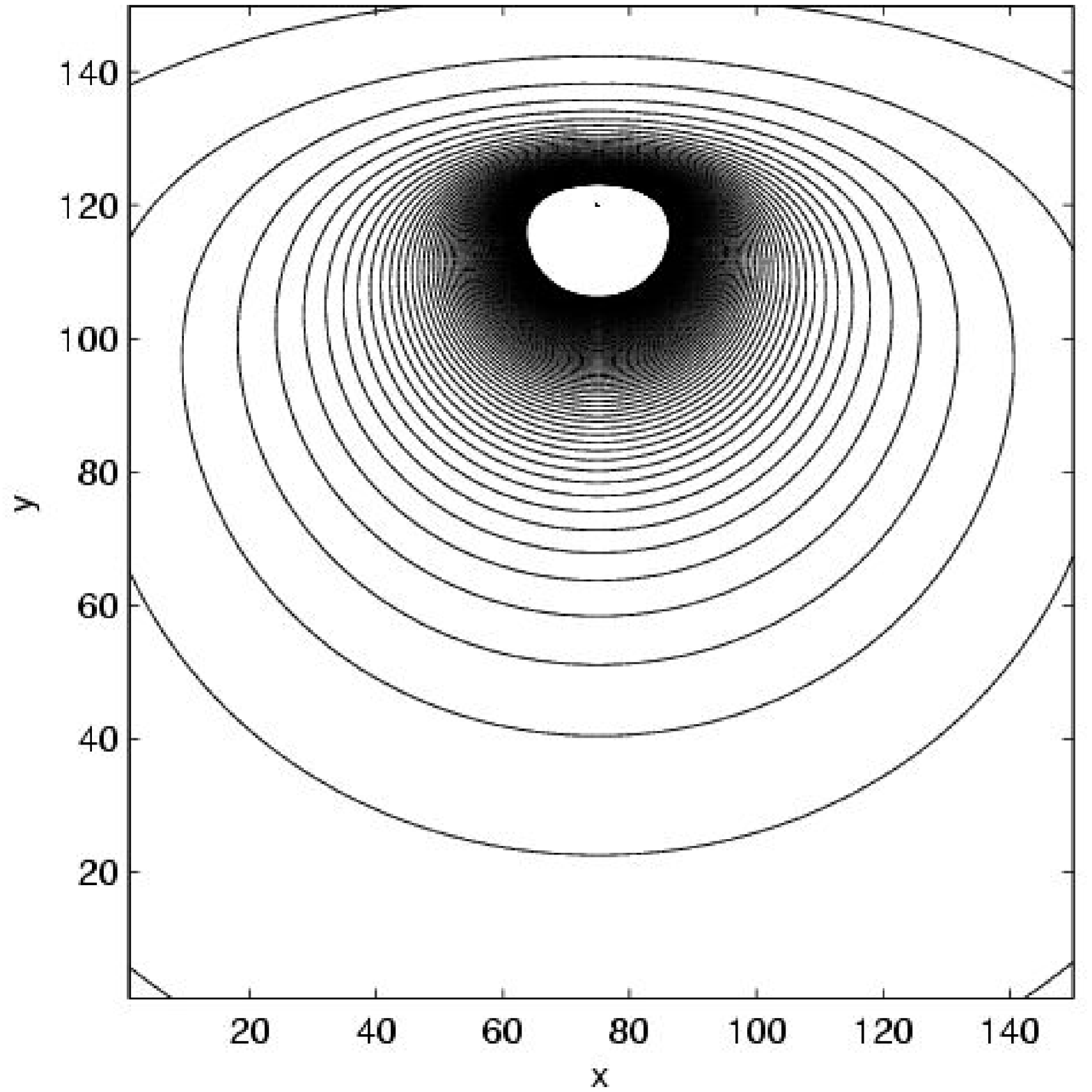}
\caption{Field of a charged particle in a homogeneous medium;
\protect{\\*}$n=2$, $v=0.45\,c$\label{simpleplot1}}
\end{figure}

\begin{figure}
\includegraphics[width=0.9\columnwidth]{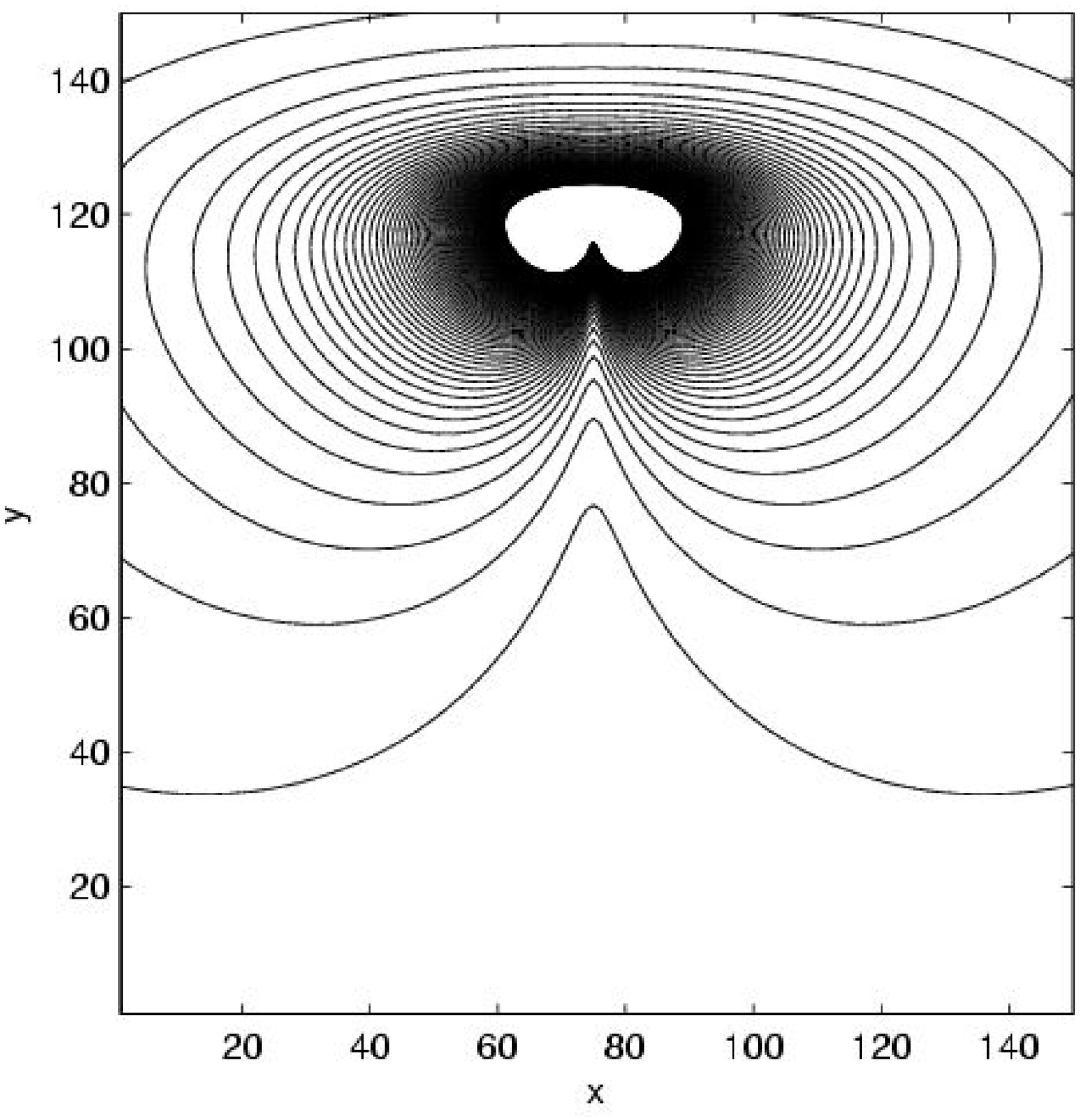}
\caption{Field of a charged particle in a homogeneous medium;
\protect{\\*}$n=1$, $v=0.9\,c$\label{simpleplot2}}
\end{figure}

\begin{figure}
\includegraphics[width=0.9\columnwidth]{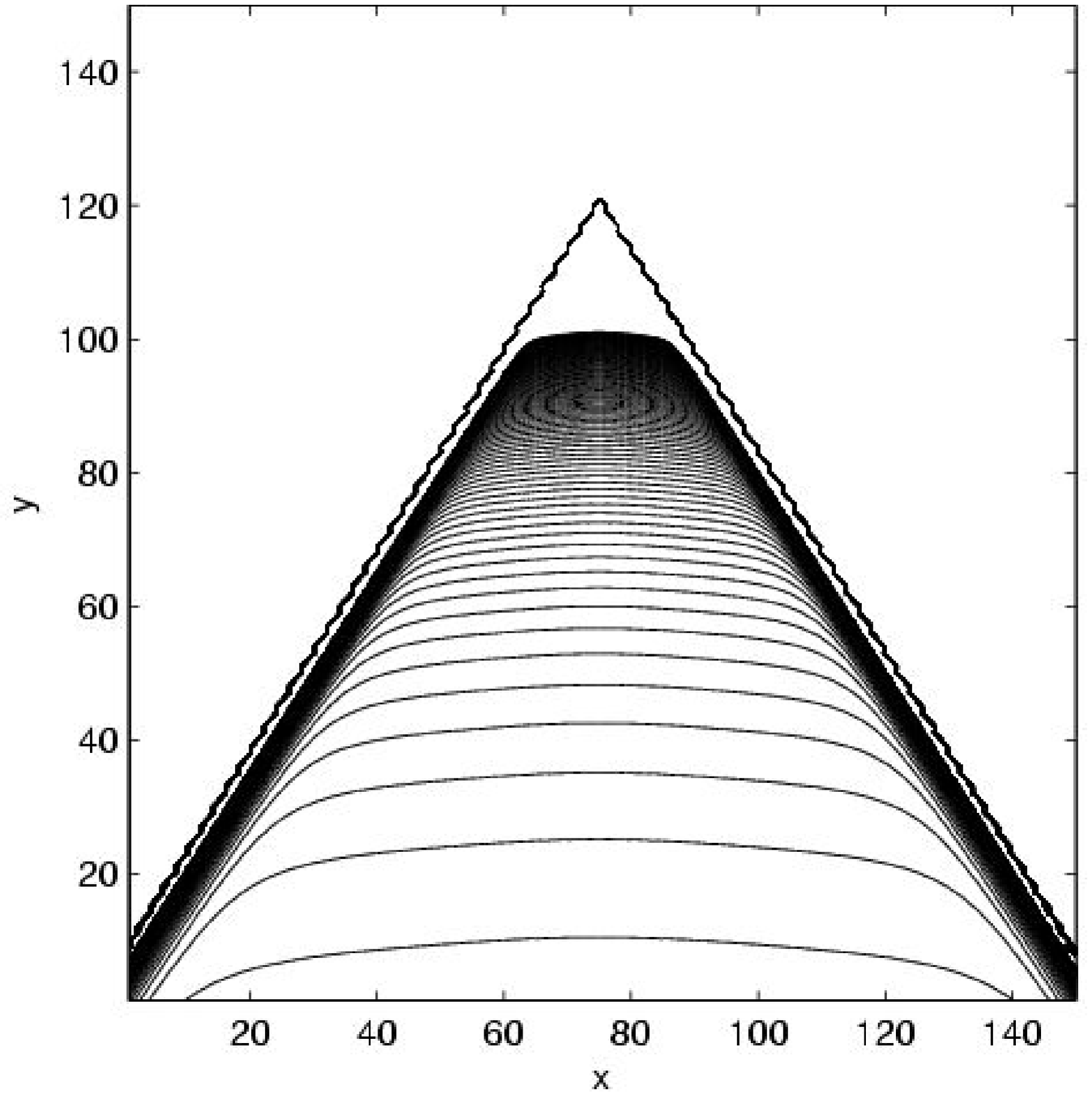}
\caption{Field of a charged particle in a homogeneous medium; 
\protect{\\*}$n=2$, $v=0.9\,c$\label{simpleplot3}}
\end{figure}

Having identified the space--time co--ordinates of our retarded source(s), we can now calculate the field at $(x^\prime,y^\prime,0,t^\prime)$ using \eref{dfield}. When this process is repeated for multiple sets of co--ordinates $(x^\prime,y^\prime,0,t^\prime)$, a plot of the field may be built up. Figures~\ref{simpleplot1}--\ref{simpleplot3} demonstrate the results which may be obtained.

\section{Results}\label{results}
\begin{figure}
\includegraphics{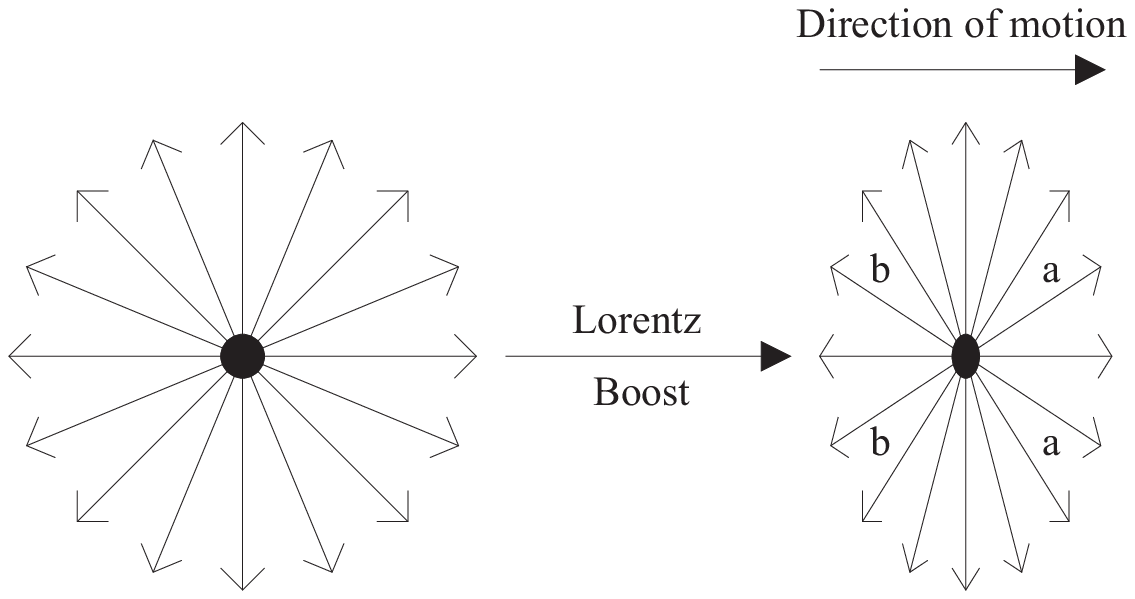}
\caption{Field line representation of a coulomb field undergoing a Lorentz boost. The particle is travelling left--to--right so regions marked `a' lie ahead of the moving particle with respect to the direction of motion, and those marked `b' lie behind it.\label{coulombboost}}
\end{figure}

To represent the field of the particle, we have plotted contours of equal $\bi{D}$ field magnitude. This should not be confused with the plotting of electric or magnetic field lines. The reader may be familiar with the representation of a Coulomb field undergoing a Lorentz boost shown in \fref{coulombboost}.
Why, then, is the field plotted in figures \ref{simpleplot1} and \ref{simpleplot2} not similarly symmetrical?

When the charge is placed in motion, it constitutes a finite current element and will therefore generate a magnetic field. The charge is moving, and hence distances to this current element will vary with time. As they do, the magnetic field will also vary with time, inducing a further electric field. This field is opposed to the existing compressed Coulomb field in the regions marked `a' in \fref{coulombboost}, and complements it in the regions marked `b', giving rise to the observed distortion in the field contours. The magnitude of this distortion is dependent on electromagnetic induction in accordance with Maxwell's equations, and hence depends on the speed of the particle relative to the speed of light \emph{in vacuo}. Hence less distortion is noted in \fref{simpleplot1} than in \fref{simpleplot2}, despite the particle's speed being 0.9 times the speed of light in the local medium in each case.



\begin{figure}
\includegraphics{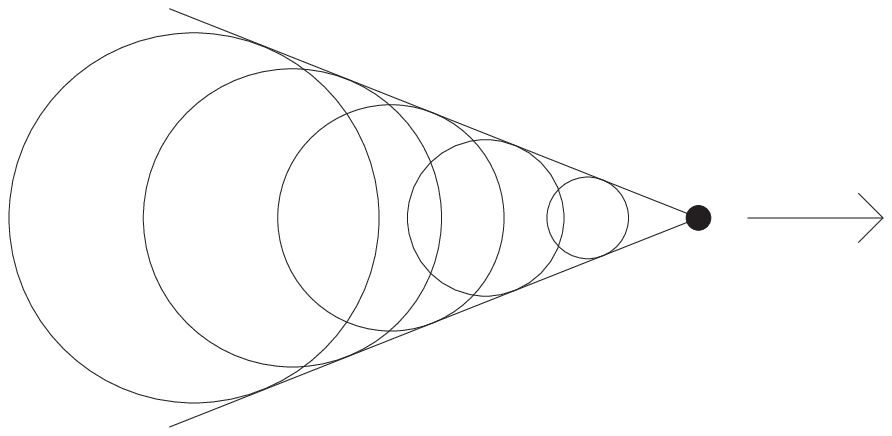}
\caption{Formation of a shock cone from repeated emission of spherical wavefronts\label{shockcone}}
\end{figure}

In \fref{simpleplot3}, the particle is now travelling superluminally and as a result continually overtakes the leading edge of its propagating field. Because of this, a shock front is built up. It is this which is perceived as the \v{C}erenkov radiation (see \sref{discussion}). An analogy which is often employed to illustrate this phenomenon is to imagine that as it travels, the particle emits repeated pulses of electromagnetic radiation, expanding in shells as shown in \fref{shockcone}.
While a useful aid to visualisation, this analogy breaks down when applied to subluminal particles and regions lying within the \v{C}erenkov cone. A more complete explanation could be developed in which the field constitutes the emission of \emph{virtual} photons in analogy with quantum field theory, but lies outside the scope of this article.

It is also interesting to see how these results relate to the time-reversal symmetry of the Maxwell equations. Essentially, time reversal interchanges the advanced and retarded potentials, and hence appropriately reverses the direction of the \v{C}erenkov cone, as would be expected if the direction of motion of the charge were reversed. This can be contrasted with the Lorentz-contracted field lines of \fref{coulombboost},
which are intrinsically symmetric, and for which time reversal simply reverses the direction of travel of the particle leaving the field lines unchanged.




\section{Discussion}\label{discussion}

\subsection{Electromagnetic Radiation}

\begin{figure}
\includegraphics[width=0.9\columnwidth]{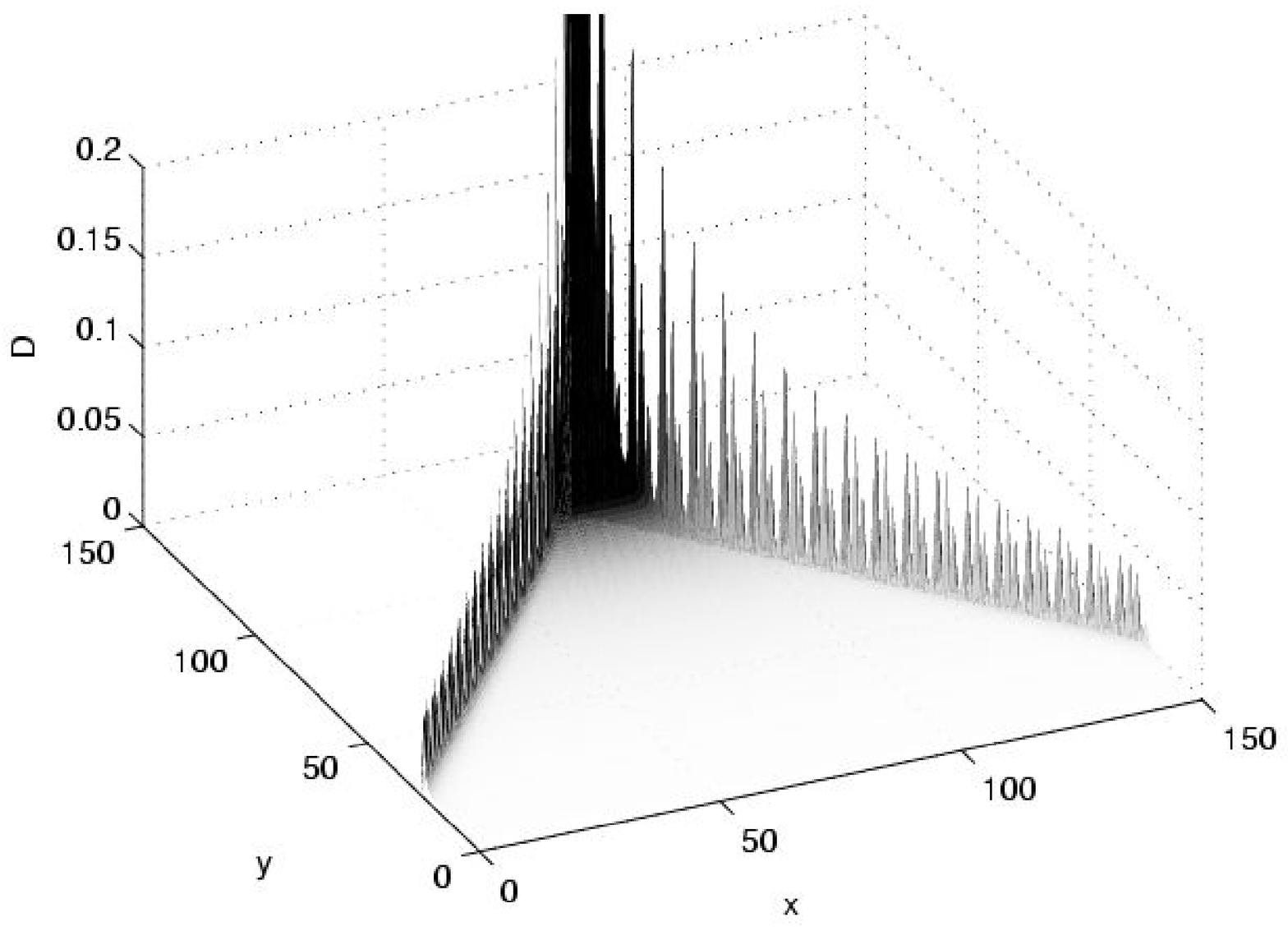}
\caption{'$\bi{D}$' field of a superluminal particle (see also \fref{simpleplot3})\label{cherenkov3d}}
\end{figure}

As is clearly shown in \fref{cherenkov3d},
the shock cone constitutes a narrow region of comparatively powerful electric and magnetic fields, giving rise to local concentrations in field energy. These regions of concentrated field propagate outward with time. Of course, a freely propagating energy-carrying wave in the electric and magnetic fields is what we know as electromagnetic radiation, in this case visible light.

In allowing the particle to continue to pursue a constant velocity trajectory in our calculations, we have neglected the effects of this radiative energy loss.

\subsection{Freedom to Propagate}

What constitutes a freely propagating electromagnetic wave? In the above section we identified the field surge of the \v{C}erenkov cone with electromagnetic radiation. But the method used to generate these images makes no allowance for free packets of fields propagating through space --- the fields involved all originate directly from the moving charge. Is it therefore appropriate to think of this wave as free in the same sense as we think of photons being free?

The answer is yes. Although in particulate models photons are considered as independent entities, and likewise in classical electromagnetics we often consider sourceless, freely propagating plane waves, in practice there exists a charge at the end of every photon or electromagnetic wave, of whose retarded fields it is in fact an extension. This charge may be accelerating, jumping between atomic orbitals, or travelling through an optically dense medium as seen here. Our `free wave' is indeed free, in that its nature and behaviour are unaffected by any subsequent actions of the originating charge: That modulation in the local electromagnetic field will continue to propagate out indefinitely at the local speed of light, even if the originating particle is subsequently somehow destroyed.

\subsection{Acoustic Shockwaves}

\begin{figure}
\begin{center}
\includegraphics[width=10cm]{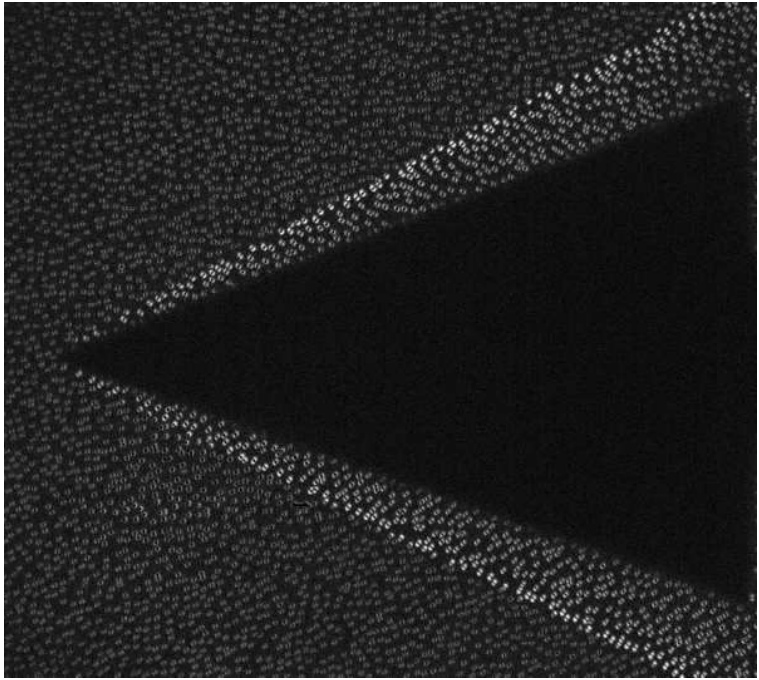}
\end{center}
\caption{Supersonic gas flow over a stationary cone at Mach 4, visualised using the background oriented schlieren technique. 
\label{figure8}}
\end{figure}

As we have seen in \sref{results}, \v{C}erenkov radiation arises due to the formation of a shockwave in a particle's electric and magnetic fields. The analogous problem in acoustics is the formation of shockwaves due to supersonic gas flows over material bodies, which continues to be of great importance in aeronautical engineering. In 1886 the first photographs of the bow shock of a supersonic projectile were created by the collaboration of Mach, Salcher and Riegler \cite{reichenbach}, utilising a technique derived by Toepler in 1864, known as the schlieren method. Related techniques continue to be in use to this day (\fref{figure8}).

Once again, the shock cone may be considered to be built up by superposition of consecutive spherical wavefronts emitted by the source as it travels (\fref{shockcone}). In this case, the wavefronts are pressure waves within the surrounding medium.
Diagrams showing the construction of the shock cone in this manner were first published by Christian Doppler in 1842 \cite{rott}, though it is after Mach that the cone is usually named, in recognition of his later experimental work.

Introduction of factors such as viscosity and turbulent flow lead to additional behaviours not discussed in this article, and cause the study of fluid dynamics and acoustic shockwaves to be a complex and fascinating field.

\section{Supplementary Material}

The electronic version of this article is accompanied by the MATLAB program \verb+Cherenkov.m+, which was used to generate the plots accompanying this article. Usage instructions may be viewed by typing `\verb+help Cherenkov.m+'.

\section{Conclusion}

This paper has aimed to illustrate the behaviours of the fields of a moving charge in an optically dense medium, including \v{C}erenkov radiation. These behaviours are readily simulated on a modern desktop computer, and demonstrate how the radiation of the \v{C}erenkov cone arises naturally from the fields of a superluminal charge. The concept of a `free' photon is discussed, in relation to its origin in the retarded field of an electric charge. It is explained how the existence of the photon arises as a result of the motion of the charge, but that the subsequent behaviours of the charge and the photon, or field wave packet, are independent. Finally, an analogy is drawn between the formation of the \v{C}erenkov cone and the formation of the bow shock of a supersonic projectile, a topic of vital importance and ongoing research in aeronautical engineering.


It is hoped that this exhibition of the interesting phenomenon of \v{C}erenkov radiation may stimulate the student to further self-guided learning, whether by developing upon the theme of this article (for example, by simulating the fields of a particle pursuing an arbitrary path, or adjacent to a medium of differing refractive index) or by investigating other specific radiative phenomena. For example, the Heaviside--Feynman equations may be used to calculate the fields of an accelerating charge as found within a radiating dipole antenna, or a synchrotron. An advanced student pursuing an interest in astrophysics may wish to investigate the radiative consequences of a straight particle path in curved space--time. There are many more interesting possibilities to explore.

\ack

We would like to thank D. Ramanah and A. Prof. D.~J. Mee at the Centre for Hypersonics, Department of Mechanical Engineering, The University of Queensland, Australia, for the image used in \fref{figure8}.

\section*{References}
\bibliographystyle{unsrt}
\bibliography{cherenkov}

\newpage
\section*{Software}
\subsection*{Cherenkov.m}
{\footnotesize \input{Cherenkov.m.tex}}

\medskip
\subsection*{ddt.m}
{\footnotesize \input{ddt.m.tex}}

\end{document}

%% file: Cherenkov.m.tex
\begin{verbatim}
function Cherenkov(ymax,yfin,tfin,v,refindex,t,plotoptions)
% Cherenkov(ymax,yfin,tfin,v,refindex,t,plotoptions)
% Calculates Maxwell's D field for a charged particle travelling through a
% homogenous medium.
% ymax sets the size of the viewing area.
% Particle travels along line y=ymax/2, reaching yfin at time tfin.
% Velocity v is a decimal multiplier of c.
% Plot is generated for the field at time t.
%
% plotoptions:
% 1: 3D
% 2: coloured surface
% 4: contour (slow)
% Options may be combined using addition.
% Default value: 7 (all).
%
% Examples:
% Try Cherenkov(150,120,100,0.8,2,100,3)
% and Cherenkov(150,120,100,0.8,1,100).

if (nargin==6)
    plotoptions=7;
end
if (nargin<6)
    'Insufficient arguments supplied. Please type help CherenkovP for more information.'
    return
end

warning off MATLAB:divideByZero

particlepos=ymax/2; % path goes along this x line
advret=+1; % +1 or -1 for advance or retarded potentials
yzero=yfin-v*tfin; %particle position at time t=0

% Need to tabulate unit R vector, R magnitude and lookback time
% for all 4-co-ords wrt particle track.

R=zeros(ymax,ymax,3,3); % x,y,t,values
Rvec=zeros(ymax,ymax,3,2,2); % x & y unit vectors - 2 sets of solutions.
% First three indices indicate location at which solutions apply. Fourth
% index indicates x and y components of solution. Fifth index indicates
% first and second valid solutions. Y solution space is also used as
% temporary workspace.

% Fill R(,1) with x positions, R(,2) with y positions, R(,3) with z
% positions, R(,4) with t positions; i.e. make R the position vector array.
R(1:ymax,1,1,1)=1:ymax;
R(1,1:ymax,1,2)=(1:ymax)';
for q=2:ymax
    R(:,q,1,1)=R(:,1,1,1);
    R(q,:,1,2)=R(1,:,1,2);
end
R(:,:,2,1:2)=R(:,:,1,1:2);
R(:,:,3,1:2)=R(:,:,1,1:2);
R(:,:,1,3)=t-1;
R(:,:,2,3)=t;
R(:,:,3,3)=t+1;

% Put working value 1 into Rvec(,1,1)
Rvec(:,:,:,1,1)= (yzero + R(:,:,:,3)*v - v.^2.*R(:,:,:,2).*refindex.^2);
Rvec(:,:,:,1,1)= Rvec(:,:,:,1,1) ./ (v.^2.*refindex.^2 - 1);

% Put working value 2 into Rvec(,1,2)
Rvec(:,:,:,1,2)= refindex.^2.*(R(:,:,:,2).^2+(particlepos-R(:,:,:,1)).^2) - R(:,:,:,3).^2;
Rvec(:,:,:,1,2)= v.^2.*Rvec(:,:,:,1,2) - yzero.^2 - 2*yzero*v*R(:,:,:,3);
Rvec(:,:,:,1,2)= Rvec(:,:,:,1,2) ./ (v.^2.*refindex.^2 - 1);

% Put source Y solution 1 into Rvec(,2,1)
Rvec(:,:,:,2,1)= -Rvec(:,:,:,1,1) + sqrt( Rvec(:,:,:,1,1).^2 - Rvec(:,:,:,1,2) );

% Put source Y solution 2 into Rvec(,2,2)
Rvec(:,:,:,2,2)= -Rvec(:,:,:,1,1) - sqrt( Rvec(:,:,:,1,1).^2 - Rvec(:,:,:,1,2) );

% Put retarded source time into sourcet
sourcet(:,:,:,1,:)=(Rvec(:,:,:,2,:)-yzero)/v;

% Change from source absolute Y to relative Y
Rvec(:,:,:,2,1) = R(:,:,:,2) - Rvec(:,:,:,2,1);
Rvec(:,:,:,2,2) = R(:,:,:,2) - Rvec(:,:,:,2,2);
% e.g. +20 means you are 20 ahead of your source point

% Store retarded source X displacement in Rvec(,1,:)
Rvec(:,:,:,1,1) = R(:,:,:,1)-particlepos;
Rvec(:,:,:,1,2) = R(:,:,:,1)-particlepos;

%sourcey=Rvec(:,:,:,1,:);

% R is now going to be displacement magnitude to retarded source:
clear R
R=zeros(ymax,ymax,3,2,2);
% Put distance from retarded source into R(,1) and R(,2):
R(:,:,:,1,:)=sqrt(sum(Rvec.^2,4));
R(:,:,:,2,:)=R(:,:,:,1,:);

% Divide Rvec by R to make it a unit vector
Rvec=Rvec./R;

% Create array for Rvec on R^2 term
RonRsq=Rvec./R./R;

'R matrices done'

Dfield=RonRsq+ddt(RonRsq).*R+ddt(ddt(Rvec));

% Eliminate source distances of zero (which create NaN entries which are
% most troublesome)

for xx=1:ymax
    for yy=1:ymax
        for sol=1:2
            for coord=1:2
                if isnan(Dfield(xx,yy,2,sol,coord))
                    Dfield(xx,yy,2,sol,coord)=0;
                end
            end
        end
    end
end

'D field done'

% Reduce array of vector D field to array of scalar magnitude of D field
Dfield=sqrt(sum(Dfield.^2,4));

% Eliminate invalid sources:

if (advret==1)
    % Eliminate sources in past:
    Dfield=Dfield.*(sourcet<t);
else
    % Eliminate sources in future:
    Dfield=Dfield.*(sourcet>t);
end

% Eliminate contributions due to imaginary particle times/positions(!)
Dfield=Dfield.*(imag(sourcet)==0);

'Unwanted sources eliminated'

% Combine valid sources by linear superposition
Dfield=sum(Dfield,5);

% Orient graph correctly for plotting
Dfield(:,:,2,:)=Dfield(:,:,2,:)';

if (bitand(plotoptions,4)) % Contour plot
    'Creating contour plot...'
    figure(3)
    v=[0:0.0001:0.01];
    contour(real(Dfield(:,:,2,:)),v);
    colormap([0 0 0]);
    xlabel('x')
    ylabel('y')
end
    
if (bitand(plotoptions,1)) % 3D plot
    'Creating surface plot...'
    figure(1)
    surfc(real(Dfield(:,:,2,:)));
    caxis([0 0.02]);
    zlim([0 0.2]);
    xlabel('x')
    ylabel('y')
    zlabel('D')
end

if (bitand(plotoptions,2)) % Flat colour plot
    'Creating colour plot...'
    figure(2)
    pcolor(real(Dfield(:,:,2,:)));
    shading flat
    caxis([0 0.02]);
    xlabel('x')
    ylabel('y')
end
\end{verbatim}

%% file: ddt.m.tex
\begin{verbatim}
function B=ddt(A)
% Takes derivative wrt time of field

[x,y,t,s,d]=size(A);
BiggerA=zeros(x,y,t+2,s,d);
B=zeros(x,y,t,s,d);
BiggerA(1:x,1:y,2:t+1,:,:)=A(1:x,1:y,1:t,:,:);

%Linearly extrapolate beyond boundaries of original A to first order

BiggerA(1:x,1:y,1,:,:)=A(1:x,1:y,1,:,:).*2-A(1:x,1:y,2,:,:);
BiggerA(1:x,1:y,t+2,:,:)=A(1:x,1:y,t,:,:).*2-A(1:x,1:y,t-1,:,:);

%Calculate ddt's

B(1:x,1:y,1:t,:,:)=( BiggerA(1:x,1:y,3:t+2,:,:)-BiggerA(1:x,1:y,1:t,:,:) )./2;
\end{verbatim}